\documentclass[aip,jcp,amsmath,amssymb,reprint]{revtex4-1} 

\usepackage{graphicx}
\usepackage{dcolumn}
\usepackage{bm}
\usepackage{caption}
\usepackage{mhchem}
\graphicspath{{./figures/}}
\begin{document}


\title[]{Stochastic Coupled Cluster Theory: Efficient Sampling of the Coupled Cluster Expansion\\}

\author{Charles J. C. Scott}
 \email{cs675@cam.ac.uk}
\author{Alex J. W. Thom}%

\affiliation{ 
Department of Chemistry, University of Cambridge, Cambridge, UK
}%

\date{\today}

\begin{abstract}
We consider the sampling of the coupled cluster expansion within stochastic coupled cluster theory. Observing the limitations of previous approaches due to the inherently non-linear behaviour of a coupled cluster wavefunction representation we propose new approaches based upon an intuitive, well-defined condition for sampling weights and on sampling the expansion in cluster operators of different excitation levels. We term these modifications even and truncated selection, respectively. Utilising both approaches demonstrates dramatically improved calculation stability as well as reduced computational and memory costs. These modifications are particularly effective at higher truncation levels owing to the large number of terms within the cluster expansion that can be neglected, as demonstrated by the reduction of the number of terms to be sampled at the level of CCSDT by 77\% and at CCSDTQ56 by 98\%.

This article may be downloaded for personal use only. Any other use requires prior permission of the author and AIP Publishing.

The following article has been accepted by the Journal of Chemical Physics. After it is published, it will be found at http://aip.scitation.org/journal/jcp/.
\end{abstract}

\maketitle

\section{Introduction}

Coupled cluster theory has been a workhorse of electronic structure theory\cite{Helgaker2008,Paldus1999,Crawford2000,Bartlett2007} since its introduction,\citep{Coester1958,Coester1960,Cizek1966,Cizek1971,Paldus1972} providing an accurate, size consistent, polynomial-scaling approximation suitable for a wide range of weakly correlated systems of interest for chemists, such as organic molecules around their equilibrium bond length.\cite{Bartlett2007} In particular the ``gold-standard'' CCSD(T) approach\citep{Raghavachari1989} provides excellent treatment of many chemically interesting problems.

However, the scaling of CCSD(T) as $\mathcal{O}$($N^7$) with system size has precluded widespread application to larger problems, and in general the high-order polynomial scaling of all truncations within canonical coupled cluster theory preclude many large-scale applications. This is exacerbated when looking beyond familiar chemical systems at equilibrium to more strongly statically correlated problems, where a high accuracy treatment demands much higher truncation levels\cite{Chan2004} and correspondingly much more expensive calculations. 

Various approaches have been developed to remedy these issues. The use of local approximations has enabled coupled cluster calculations of various forms to be performed with computational cost scaling approximately linearly with system size\cite{Schutz2001,Schutz2002,Schutz2003,Flocke2004,Subotnik2006,Ziokowski2010,Riplinger2013a,Riplinger2016,Hughes2008,Riplinger2013,Li2006,Li2009} while retaining relatively controlled errors compared to canonical results. 
Various multireference coupled cluster ansatzes also allow accurate treatment of strong correlation\cite{Pal1987,Oliphant1991,Piecuch1993,Mahapatra1998,Mahapatra1999,Lyakh2012} without dramatically increased truncation levels, while various approaches allow access to these higher truncation levels without prohibitive costs.\cite{Shen2012,Bauman2017}

In the same time period, various other approaches to the solution of the electronic Schr\"odinger equation have been developed. Projector Monte Carlo methods, based around the stochastic propagation of a trial function form to the ground state wavefunction via application of the Hamiltonian, have proved particularly effective.
Diffusion Monte Carlo (DMC) has shown great utility, providing variational energy estimates and energy differences to within chemical accuracy (1kcal $\textrm{mol}^{-1}$) in a wide variety of systems.\citep{Foulkes2001} Its scaling as $\mathcal{O}$($N^3$) with system size\cite{Foulkes2001} and almost perfect embarrassing parallelism enables relatively straightforward application to large systems, with a number of high-quality codes available.\citep{Wagner2009,Needs2010a,Kim2012} Its main limitation is the fixed-node approximation which leads to an uncontrolled error within the resulting calculations.\citep{Reynolds1982} A variety of trial function forms\cite{Filippi1996,Bajdich2006a,Bajdich2008} and backflow transformations\citep{L??pezR??os2006} have been used to reduce this error, although the trade-off between the increased computational cost associated with evaluating a more complicated wavefunction form must be weighed against the likely energy improvement given the variational nature of the method.

Further progress has been made by combining stochastic propagation with traditional quantum chemical approaches to representing electronic wavefunctions, namely via a combination of Slater determinants. Work by Booth \textit{et al.} combining PMC methods with Full Configuration Interaction (FCI) resulted in the Full Configuration Interaction Quantum Monte Carlo (FCIQMC) method\citep{Booth2009} that is perhaps the most well-studied of the resulting methods.

While still exponentially scaling in memory and computational cost FCIQMC has a dramatically reduced prefactor compared to FCI and does not suffer from any uncontrollable approximations. For mid-sized systems where relativistic effects are negligible it can thus provide a variationally exact energy estimate without requiring huge amounts of memory or computational time. The initiator adaptation\citep{Cleland2010}(\textit{i}-FCIQMC) of this method provides a lowered exponential exponent,\citep{Cleland2011} albeit while requiring convergence of the initiator error with increasing population.

\textit{i}-FCIQMC has proved capable of treating a wide range of interesting systems,\citep{Booth2011,Shepherd2012a,Booth2013,Sharma2014a} while also being extended to obtain system properties,\citep{Overy2014,Thomas2015} treat excited states,\citep{Blunt2015b,Blunt2017} capture static correlation from a larger space\citep{Manni2016,Thomas2015b} and use explicitly correlated approaches to avoid explicit extrapolation to the complete basis set limit.\citep{Kersten2016} With appropriate implementation it can also effectively leverage modern computing resources\citep{Booth2014} via massive parallelism. However, its extension to larger systems is still precluded by its scaling with system size, although recent modifications to apply a similar approach to nonlinear wavefunctions improve this issue.\cite{Schwarz2017}

Previous work\citep{Thom2010,Spencer2015a} has demonstrated the successful combination of the CC and PMC approaches to give a stochastic solution to the coupled cluster equations, reproducing deterministic results within stochastic error bars while storing only a fraction of the total available Hilbert space. This approach is termed Coupled Cluster Monte Carlo (CCMC).

Within this paper we will first give an overview of the CCMC method in section \ref{sec:CCMC} then look specifically at the approaches taken to selecting terms within the coupled cluster expansion in section \ref{sec:old-selection}. Based on considerations from these previous methods we then propose a new approach to selection in section \ref{sec:new-selection}, before presenting results obtained using this approach in section \ref{sec:results}.

\section{Coupled Cluster Monte Carlo}
\label{sec:CCMC}

We first detail the approach of CCMC without specifying the means of selection from the coupled cluster expansion. Full details of and background to the algorithm can be found elsewhere.\cite{Thom2010,Spencer2015a}

As for all projector methods we start by applying a Wick rotation to the time-dependent Schr\"{o}dinger equation, then solving the resulting differential equation to obtain the imaginary-time dynamics as $|\Psi(\tau)\rangle \propto e^{-\tau\hat{H}}|\Psi(\tau=0)\rangle$. Performing an eigendecomposition of the initial wavefunction $|\Psi(\tau=0)\rangle$ and assuming it has nonzero overlap with the ground state of the Hamiltonian $|\Psi_0\rangle$ we obtain the expression

\begin{equation}
|\Psi_0\rangle \propto \lim\limits_{\tau\to\infty} e^{-\tau \hat{H}} |\Psi(\tau=0)\rangle.
\end{equation}
If we set the constant of proportionality equal to $e^{\tau E_0}$ to maintain a constant non-zero normalisation in the limit $\tau\to\infty$ this becomes

\begin{equation}
|\Psi_0\rangle = \lim\limits_{\tau\to\infty} e^{-\tau (\hat{H} - E_0)} |\Psi(\tau=0)\rangle
\label{eq:exponential-propagation}
\end{equation}
assuming that our initial wavefunction was normalised to some condition. 

Approximating the exponential propagator by repeated application of the linear propagator $1 - \delta\tau (\hat{H} - E_0)$ gives
\begin{equation}
|\Psi_0\rangle = \lim\limits_{N\to\infty} \left[ 1 - \delta\tau (\hat{H} - E_0)\right] ^ N |\Psi(\tau=0)\rangle.
\label{eq:linear-propagation}
\end{equation}
It should be noted that equality still holds in Eq. \eqref{eq:linear-propagation}, since the exponential and linear propagators have identical eigenfunctions and so no timestep error is introduced. This does require that the timestep $\delta\tau$ is smaller than the spectral range of the Hamiltonian.\cite{Booth2009}

To project out the ground state wavefunction we thus require the ability to calculate $|\Psi(\tau + \delta\tau)\rangle$ from arbitrary $|\Psi(\tau)\rangle$ using the relation

\begin{equation}
|\Psi(\tau + \delta\tau)\rangle = \left[ 1 - \delta\tau (\hat{H} - S)\right] |\Psi(\tau)\rangle.
\label{eq:generic-propogation}
\end{equation}
Within this expression the unknown ground state energy $E_0$ has been replaced by an arbitrary constant $S$ that determines the rate and direction of change in magnitude of $|\Psi(\tau)\rangle$.

Within CCMC we combine this approach with the Coupled Cluster wavefunction ansatz\citep{Thom2010}

\begin{equation}
\Psi_\textrm{CCMC}(\tau) = N_0(\tau)e^{\frac{\textit{\^{T}}(\tau)}{N_0(\tau)}}|D_0\rangle,
\label{eq:CC-ansatz}
\end{equation}
where $\hat{T}$ is the cluster operator

\begin{equation}
\hat{T}(\tau) = \hat{T}_1(\tau) + \hat{T}_2(\tau) + \hat{T}_3(\tau) + \ldots,
\end{equation}
and $\hat{T}_i$ contains operators creating all $i$-orbital replacements of the reference determinant, in the form of excitors $\hat{a}_{\textbf{i}}$. These excitors are combinations of second-quantised creation and annihilation operators, such that their action upon a given reference determinant results in the Slater determinant with the corresponding orbital replacements

\begin{equation}
T_i(\tau) = \sum_{\textbf{j}\in i^{\textrm{th}} \textrm{replacements}} t_{\textbf{j}}(\tau) \hat{a}_{\textbf{j}}.
\end{equation}

This wavefunction form has the advantage of being size-consistent with a truncation of $\hat{T}(\tau)$ at any excitation level, for instance $\hat{T}(\tau) = \hat{T}_1(\tau) + \hat{T}_2(\tau)$ gives the coupled cluster singles and doubles (CCSD) wavefunction form. This enables consideration of a much smaller polynomial-scaling portion of the space compared to methods such as FCIQMC, which require consideration of the full (exponentially scaling) Hilbert space to remain size-consistent.

To utilise this ansatz in combination with \eqref{eq:generic-propogation} we must substitute this form and project onto a given determinant $\langle D_{\textbf{n}} |$. Representing this via the action of the corresponding excitor avoids further complications due to sign changes resulting from the interchange of second quantised operators. We thus obtain

\begin{equation}
\langle D_0 | \hat{a}_{\textbf{n}}^\dagger|\Psi_\textrm{CCMC}\rangle \rightarrow
\langle D_0 | \hat{a}_{\textbf{n}}^\dagger[ 1 - \delta\tau (\hat{H} - S)] |\Psi_{\textrm{CCMC}}\rangle.
\label{eq:scct-propogation}
\end{equation}

Noting that $\langle D_0 | \hat{a}_{\textbf{n}}^\dagger|\Psi_\textrm{CCMC}\rangle = t_{\textbf{n}} + \mathcal{O}(\hat{T}^2)$ and cancelling higher order terms on both sides we obtain

\begin{equation}
t_{\textbf{n}} \rightarrow t_{\textbf{n}} - \delta \tau\langle D_0 | \hat{a}_{\textbf{n}}^\dagger [\hat{H}-S]|\Psi_{\textrm{CCMC}}\rangle.
\label{eq:scct-propogation2}
\end{equation}

This gives us an iterative procedure for solving the coupled cluster equations to an arbitrary truncation level.\footnote{Following previous work,\cite{Franklin2016} to ensure convergence when $S$ does not average to the coupled cluster energy these dynamics are modified to include the projected energy explicitly, though this does not affect the sampling of $\Psi_{CC}$.} 

Reductions in storage costs are achieved by representing the values $t_{\textbf{i}}$ by a population of discretised excitor particles (excips) and interpreting the iterative procedure in \eqref{eq:generic-propogation} as a population dynamic.

This requires sampling the action of the Hamiltonian upon our current population distribution. We observe that expanding the full instantaneous CCMC wavefunction is prohibitively expensive, due to the large number of relatively negligible terms resulting from larger cluster sizes. We instead sample terms from within it and reweight for their selection probability.

This will be expounded in more detail in the next section of this paper, but for now we will simply assume we have generated with a normalised probability $p_{\textrm{select}}(e)$ a combination of excitors $e$ with total amplitude $w_e$. When applied to the reference this collection of excitors collapses to give $w_e |D_\textbf{m}\rangle$. If we are making $n_{\textrm{a}}$ such generation attempts on a given iteration our expression becomes

\begin{equation}
t_{\textbf{n}} \rightarrow t_{\textbf{n}} - \delta \tau\langle D_0 | a_{\textbf{n}}^\dagger [\hat{H}-S] \sum\limits_{e}^{n_{\rm{a}}} \frac{w_{e}}{n_{\rm{a}}p_{\textrm{select}}(e)} | D_{\textbf{m}}(e)\rangle.
\end{equation}

Having selected $w_{e}|D_{\textbf{m}}(e)\rangle$ we then sample the action of the Hamiltonian on it via:

\begin{enumerate}
	\item Spawning: choose a random single or double excitation of $|D_{\textbf{m}}\rangle$ (denoted $|D_{\textbf{n}}\rangle$) and create an excip with probability 
	\begin{equation}
	p_{\textrm{spawn}} = \delta\tau\frac{|w_e|}{n_{\rm{a}}p_{\rm{select}}(e)}\frac{|H_{\textbf{nm}}|}{p_{\textrm{gen}}}
	\label{eq:pspawn}
	\end{equation}
	on the excitor $\hat{a}_{\textbf{n}}$. The sign of the spawned excips is determined by that of $H_{\textbf{nm}}$.
	\item Death: spawn a particle of opposite sign to $w_eH_\textbf{nn}$ onto excitor $\hat{a}_\textbf{n}$ with probability
	\begin{equation}
	p_{\textrm{death}} = \delta\tau\frac{|w_e|}{n_{\rm{a}}p_{\textrm{select}}(e)}|H_\textbf{nn}-S|.
	\end{equation}
\end{enumerate}

We draw the reader's attention to the presence of the factor $\frac{|w_e|}{n_{\rm{a}}p_{\textrm{select}}(e)}$ within the expressions for both $p_{\textrm{spawn}}$ and $p_{\textrm{death}}$. The importance of this will be examined later.

Following this prescription results in a population dynamic that requires a critical excip population (termed the plateau or shoulder)\citep{Thom2010,Spencer2015a,Franklin2016} to give a stable calculation, but above this population the excip distribution samples the ground state wavefunction and returns the correct energy. The plateau is conventionally the limiting factor on the size of system that can considered, so is an important measure of any algorithmic changes.

We use the form of population control previously adopted for FCIQMC\citep{Booth2010} and originally suggested for DMC,\citep{Umrigar1993c} and a projected energy estimator. This gives an energy estimate in the form

\begin{equation}
E_{\textrm{proj}} = \frac{\langle D_0 | \hat{H} | \Psi_{\textrm{CCMC}}\rangle}{\langle D_0 | \Psi_{\textrm{CCMC}} \rangle}.
\end{equation}

We can sample the numerator and denominator of this expression separately when our simulation is under population control. Performing a reblocking analysis\citep{Flyvbjerg1989} enables us to remove serial correlation from our estimates of these quantities and obtain a measure of the stochastic error in these values.

This relatively simple picture may be complicated by, for instance, using non-integer walker weights\citep{Overy2014} or the details of excitation generation within spawning,\cite{Holmes2016} but remains a general description of the approach taken.

\section{Sampling the Coupled Cluster expansion: Current Approaches}
\label{sec:old-selection}

In this section, we will detail previous approaches to the selection of the $w_{e}$ and $p_{\textrm{select}}(e)$ parameters introduced in the previous section. These chart the evolution of ideas relating to this sampling, and based upon these developments we propose a new algorithm.

It should be noted that provided the algorithm used selects all terms within the expansion below with nonzero $p_{\textrm{select}}$ the result obtained should be statistically identical within the derivation of the method already provided. It can however lead to algorithmic complications, as will be examined later, and in particular can be expected to affect the plateau and so calculation computational and memory costs.

First, we will go through a few useful definitions. We wish to sample terms from within

\begin{align}
|\Psi_{\textrm{CCMC}}\rangle &= N_0 e^{\frac{\hat{T}}{N_0}} | D_0\rangle\\
	&= \left(N_0 + \sum_{\textbf{i}} t_{\textbf{i}} \hat{a}_{\textbf{i}} +  \sum_{\textbf{i,j}} \frac{t_{\textbf{i}} t_{\textbf{j}} \hat{a}_{\textbf{i}}\hat{a}_{\textbf{j}}}{N_02!} + \cdots \right) |D_0\rangle.
	\label{eq:orig-cc-expansion}
\end{align}
This is an expansion in terms containing increasing numbers of excitors. As in previous work these terms are referred to as clusters, and a cluster includes both the constituent excitors and any coefficient contributions. Evaluating the action of a cluster upon the reference is referred to as collapsing the cluster, and the number of excitors $s$ within a cluster is referred to as its size. Clusters with $s=1$ are termed non-composite and clusters with $s > 1$ are termed composite.

We will refer to the population of excips on an excitor \textbf{i} as $N_{\textbf{i}}$, and the total absolute excip population as $N_{\textrm{ex}} = \sum_{\textbf{i}} |N_{\textbf{i}}|$. In addition, we will refer to clusters and excitors by the excitation level of the determinant they result in when applied to the reference determinant. For instance, we would refer to an excitor (cluster) containing a single creation and annhiliation operator as an excitation level 1 excitor (cluster).

\subsection{Original Algorithm}

In previous work\citep{Thom2010} an initial algorithm was used that selected purely stochastically from within all possible clusters. This selected a given combination of excitors $e$ by

\begin{enumerate}
\item Selecting a cluster size $s$ following an exponential distribution (ie. $p_{\textrm{size}}(s) = \frac{1}{2^{s+1}}$ ).
\item Selecting $s$ excitors from within the current distribution and reordering them to some condition with probability $p_{\textrm{clust}}(e|s) = s!\prod_{i=1}^{s} \frac{|N_i|}{N_{\textrm{ex}}}$.
\end{enumerate}

 The factor of $s!$ in $p_{\textrm{clust}}$ can be equivalently considered to either result from the number of possible ways to select the same cluster in different orderings (if we reorder the excitors to satisfy some condition) or from the coupled cluster expansion itself. 

This approach gives $p_{\textrm{select}}(e) = p_{\textrm{size}}(s)p_{\textrm{clust}}(e|s)$ for a given number of selection attempts, and can clearly generate all possible combinations of excitors.

This approach is slightly improved by observing that since $\hat{H}$ is a two-body operator and each excitor is at least a one-body operator we can limit $s$ to the range $0\le s\le l + 2£$, where $l$ is the highest term included within the cluster operator. With this modification we use $p_{\textrm{size}}(s = l + 2) = \frac{1}{2^{l+2}}$ to ensure normalisation of $p_{\textrm{size}}$, and avoid sampling larger clusters. For the continuing analysis we ignore this modification for ease of notation, as its inclusion has minimal effect on our results.

The number of total attempts to make at selecting clusters at each iteration, $n_{\textrm{a}}$, is then set equal to $N_{\textrm{ex}}$ for convenience.

This selection scheme gives the value of $p_{\textrm{select}}(e)$ for a given cluster $e$ of size $s$ as

\begin{align}
	p_{\textrm{select}}(e) &= \frac{s!\prod_{i=1}^{s}\frac{|N_i|}{N_{\textrm{ex}}}}{2^{s+1}}\\
	&=\frac{s!\prod_{i=1}^{s}|N_i|}{2^{s+1}N_{\textrm{ex}}^{s}}.
\end{align}
Observing that

\begin{equation}
w_{e} = \frac{\prod_{i=1}^{s}|N_i|}{N_0^{s-1}},
\label{eq:w-expression}
\end{equation}
we can obtain an expression for the value $\frac{w_e}{n_{\textrm{a}}p_{\textrm{select}}(e)}$ previously observed to appear within our expressions for $p_{\textrm{spawn}}$ and $p_{\textrm{death}}$

\begin{equation}
\frac{w_e}{n_ap_{\textrm{select}}(e)} = \frac{2^{s+1}}{s!} \left(\frac{N_{\textrm{ex}}}{N_0}\right)^{s-1}.
\end{equation}

Given that in a usual calculation $\frac{N_{\textrm{ex}}}{N_0} \gg 1$ and that this ratio increases with the size of the system being considered the spawning and death probabilities of large composite clusters will thus be much greater than those of smaller clusters.

\subsection{Full Non-composite}

To remedy some shortcomings of the original approach in stability a new algorithm was implemented by Spencer, Vigor and Thom. This has not been previously detailed, but will form part of a forthcoming publication. For ease of consideration we will include a brief description of its selection considerations here.

This approach utilises a FCIQMC-like deterministic selection of the reference and non-composite components of the coupled cluster expansion but continues to use the previous algorithm for the composite cluster selection. This ensures that for $s=0,1$ we must have $\frac{w_e}{p_{\textrm{select}}(e)} = 1$, while effectively requiring $N_{\textrm{ex}} + N_0$ attempts. For composite clusters we can follow the previous prescription.

We make $N_{\textrm{ex}}$ selections of composite clusters with $p_{\textrm{size}}(s) = \frac{1}{2^{s-1}}$ while using the same approach to $p_{\textrm{clust}}(e|s)$, so obtain

\begin{align}
	n_ap_{\textrm{select}}(e\in \textrm{composite}) &= n_{a}\frac{s!\prod_{i=1}^{s}\frac{|N_i|}{N_{\textrm{ex}}}}{2^{s-1}}\\
	&=\frac{s!\prod_{i=1}^{s}|N_i|}{2^{s-1}N_{\textrm{ex}}^{s-1}}.
\end{align}
And so overall

\begin{equation}
\frac{w_{e\in \textrm{non-composite}}}{n_{\textrm{a}}p_{\textrm{select}}(e\in \textrm{non-composite})} = 1
\end{equation}

\begin{equation}
\frac{w_{e\in \textrm{composite}}}{n_ap_{\textrm{select}}(e\in \textrm{composite})} = \frac{2^{s-1}}{s!} \left(\frac{N_{\textrm{ex}}}{N_0}\right)^{s-1}.
\end{equation}

This approach provides a much more stable calculation and energy estimate. The reason for the improved projected energy estimate stability can be seen by considering the form of the integral being sampled given that $\hat{H}$ is a two-body operator

\begin{align}
E_{\textrm{proj}} =& \frac{\langle D_{0} | \hat{H} | \Psi_{\textrm{CCMC}}\rangle}{\langle D_{0} | \Psi_{\textrm{CCMC}}\rangle}\\
=& \frac{\langle D_0 | \hat{H} (1+\hat{T}_1 + \hat{T}_2 + \hat{T}_1^2)| D_0\rangle}{N_0}.
\end{align}

Within this expression, we exactly sample the action of $\hat{T}_{1}$ and $\hat{T}_{2}$ upon the reference and the value of $N_0$, meaning that the only term in this integral that is not evaluated exactly (for our instantaneous excip distribution) is $\hat{T}_{1}^2$. This drastically decreases the variance of our energy estimate.

\section{Sampling the Coupled Cluster expansion: New Approaches}
\label{sec:new-selection}

\subsection{The Need for new Approaches}

\label{sec:justify-new-approach}

In both previous methods variation in $\frac{w_e}{p_{\textrm{select}}(e)}$ for composite clusters still results in variation in $p_{\textrm{spawn}}$ and $p_{\textrm{death}}$ between different cluster sizes by a potentially very large factor. While in theory, at large enough cluster sizes, $s!$ will grow faster than $\left(2\frac{N_{\textrm{ex}}}{N_0}\right)^{s-1}$, in practice for all but totally minuscule systems $\frac{N_{\textrm{ex}}}{N_0}$ is large enough that the larger a cluster is, the greater its value of $\frac{w_e}{p_{\textrm{select}}(e)}$.

This leads to large values of $p_{\textrm{spawn}}$ and so spawning events resulting in multiple identical excips, termed blooms. These can vary in size, but can easily result in $\mathcal{O}(10^6)$ excips being produced in a single spawning event for larger systems and cause various issues within our calculations.
In the worst case they result in such a large population on an otherwise unimportant excitor within our wavefunction representation (i.e. only weakly contributing to the ground state) that the overlap $\langle\Psi_0|\Psi(\tau)\rangle$ is reduced enough to destabilise continued projection. In milder cases this destabilising effect increases the number of occupied excitors required before the plateau is reached, reducing the size of systems that can be tenably approached using this method and requiring a smaller timestep $\delta\tau$ before a calculation is stable.

\begin{figure}
{\centering
  \includegraphics[width=0.45\textwidth]{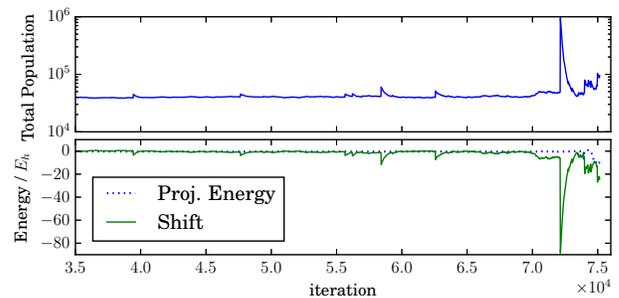}

}
\caption{Calculation dynamics for a (CCSDTQ)MC calculation on two well-separated neon atoms in a Dunning cc-pVDZ basis\citep{DunningJr1989} using full non-composite propagation. In each case the top panel shows the total population and the bottom panel shows the energy estimators.}
\label{fig:bigbloom_calculation}
\end{figure}

An example of this within a calculation is shown in Figure \ref{fig:bigbloom_calculation}, where a series of blooms, the largest of size $\approx 9 \times 10^5$ excips, occurred when the previous total population was $4\times 10^4$. This resulted in a calculation that previously appeared stable under population control becoming unstable. The blooms are visible as spikes in the total population (note the logarithmic scale) that are mirrored within the shift as the population control algorithm attempts to compensate.

Aside from the destabilising effects of these blooms, their effect upon the shift, and to a lesser extent projected energy, will result in the energy estimator being distributed highly asymmetrically about the true value and certainly not in a Gaussian distribution, as can be clearly seen within Figure \ref{fig:bigbloom_calculation}. This is a major issue as the reblocking analysis\citep{Flyvbjerg1989} usually used to obtain an estimate and errorbar for the energy assumes an approximately Gaussian distribution. Much more minor deviations from a Gaussian distribution in continuum Monte Carlo due to heavy-tailed distributions have been shown to have significant effects upon the reliability of stochastic error bars,\citep{Trail2008} thus avoidance of these effects should be an imperative to ensure reliable error bars.

Of particular concern is the dependence of $p_{\textrm{spawn}}$ on $\frac{N_{\textrm{ex}}}{N_0}$ due to two observations about the behaviour of $\frac{N_{\textrm{ex}}}{N_0}$, the `particle ratio'. Firstly, the equilibrium value of $\frac{N_{\textrm{ex}}}{N_0}$ grows with system size, truncation level and the difficulty of treating the system with the coupled cluster approximation, as represented by the contribution of the reference determinant to the ground state wavefunction. As such as we consider larger and more challenging systems our calculations will grow more difficult in an unpredictable manner.

Secondly, as previously noted,\citep{Spencer2015a} this `particle ratio' rises to a maximal value when a CCMC calculation exits the plateau due to the reference excip population previously growing more slowly than the rest of the space. This will lead to larger blooms occurring exactly when we seek to converge to a stable representation of our wavefunction, potentially delaying exit from the plateau, increasing the maximal particle ratio and exacerbating the problem.

Together, it seems likely that extension of our current approaches to more difficult systems will result in rapid destablisation of the calculation by blooming. Though the initiator approximation has been shown to improve stability the additional cost resulting from the slow convergence of the initiator error\citep{Spencer2015a} limits its usefulness in practice.

We will now attempt to propose an approach that avoids these problems and see what benefits this may provide.

\subsection{Even Selection}
\label{sec:even-selection}

Based upon our previous analysis, we can suggest the form of $p_{\textrm{size}}(s)$ for $s\geq 2$ should satisfy

\begin{equation}
\frac{w_{e\in \textrm{composite}}}{n_{\textrm{a}}p_{\textrm{size}}(s)p_{\textrm{clust}}(e|s)} = 1
\end{equation}
as full non-composite did for $s = 0,1$. By substituting the expressions for $p_{\textrm{clust}}$ and $w_{\textrm{j}}$ already provided and rearranging we can obtain the form

\begin{equation}
n_{\textrm{a}} p_{\textrm{size}}(s) = \frac{N_{\textrm{ex}}^s}{N_0^{s-1}s!}.
\end{equation}

We note that the value $n_{\textrm{a}} p_{\textrm{size}}(s)$ is in fact the number of attempts each cluster size requires to be sampled such that this condition holds, and that this expression returns our expected values when applied to clusters of size 0 and 1 ($N_0$ and $N_{\textrm{ex}}$ respectively).

We finally note that due to the condition that $p_{\textrm{size}}$ be normalised, $n_{\textrm{a}}$ must be varied depending upon our excip distribution between the reference and rest of the space. Via either the normalisation of $p_{\textrm{size}}$ or summing individual required attempts for a cluster expansion truncated at level $l$ we obtain the expression

\begin{equation}
n_{\textrm{a}}  = N_{\textrm{ex}}\sum_{s=0}^{l+2} \frac{1}{s!}\left(\frac{N_{\textrm{ex}}}{N_0}\right)^{s-1}.
\label{eq:nattempts-pure-even}
\end{equation}
This gives us a number of selection attempts that scales linearly with the total excip population (assuming particle ratio held constant) but exponentially with increasing truncation level $l$. For any systems of interest it can be assumed $\frac{N_{\textrm{ex}}}{N_0}>1$.

While hopefully providing a more stable calculation, this approach would rapidly become untenable for all but the very smallest systems and lowest truncation levels due to the inordinate number of samples required at larger cluster sizes. We will now detail the source of this rapid increase in cost and propose a modified algorithm to avoid it.

\subsection{Truncated Selection}
\label{sec:truncated-selection}

We start with the observation that for any given truncation level of CC many available combinations of excitors collapse to excitation levels greater than $l+2$ and thus cannot contribute to any stored coefficient via spawning. These selections are effectively wasted, leading to the large increase in $n_{\textrm{a}}$ observed previously not necessarily being tied to the number of particles created via spawning.

To illustrate this we consider our expansion in terms of the cluster operators $\hat{T}_1 + \hat{T}_2 + \cdots$. For CC at truncation level $l$ we observe that

\begin{align}
N_0e^{\hat{T}}|D_0\rangle &= N_0e^{\frac{\sum_{i=1}^{l}\hat{T}_i}{N_0}}|D_0\rangle\\
&= N_0\prod_{i=1}^{l}e^{\frac{\hat{T}_i}{N_0}}|D_0\rangle\\
&= N_0 \prod_{i=1}^{l}
\left(1 + \frac{\hat{T}_i}{N_0} + \frac{\hat{T}_i^2}{2! N_0^2} +\cdots\right) |D_0\rangle.
\label{eq:new-cc-expansion}
\end{align}
We note that the factorial prefactors of composite terms within this full expansion will differ compared to \eqref{eq:orig-cc-expansion}. This results from the enforced ordering of the operators $\hat{T}_i$ within this expansion.

We now consider performing an illustrative CCSD calculation and sampling $\Psi_{\textrm{CCMC}}$ with $\hat{T} = \hat{T}_1 + \hat{T}_2$. Considering the expansion \eqref{eq:new-cc-expansion} to cluster size $l$+2 (=4) we obtain the expression our previous approaches have effectively been dealing with as

\begin{align}
\label{eq:full-ccsd-combo-expansion}
N_0e^{\hat{T}}|D_0\rangle = &\left[N_0 \phantom{\hat{T}_2^4}\right. \\
& + \hat{T}_1 + \hat{T}_2 \nonumber \\
 + \frac{1}{2N_0} &\left(\hat{T}_1^2 + 2\hat{T}_1\hat{T}_2 + \hat{T}_2^2  \right) \nonumber \\
 + \frac{1}{6 N_0^2} &\left(\hat{T}_1^3 + 3 \hat{T}_1^2 T_2 + 3 \hat{T}_1 T_2^2 + \hat{T}_2^3 \right) \nonumber \\
 + \frac{1}{24 N_0 ^3} &\left( \hat{T}_1^4 + 4 \hat{T}_1^3 T_2 + 6 \hat{T}_1^2 T_2^2 + 4 \hat{T}_1 T_2^3 + \hat{T}_2^4 \right) \nonumber \\
+ \mathcal{O}(\hat{T}^5) & \hspace*{50pt}\left.\phantom{\hat{T}_2^4}\right] |D_0\rangle, \nonumber
\end{align}
where combinations of the same size $s$ have the same prefactor $\frac{1}{N_0^{s-1}}$. The computational inefficiency of this approach is particularly clear in the sampling of clusters of size 4, where only the $\hat{T}_1^4$ term can result in a cluster that can contribute to any integral $\langle D_{\textbf{i}} | \hat{H} | \Psi_{\textrm{CCMC}} \rangle$ for a stored $t_{\textbf{i}}$.

For a solution to this problem we can conveniently just consider sampling the product \eqref{eq:new-cc-expansion} instead of the full exponential coupled cluster expansion \eqref{eq:orig-cc-expansion}. Our size of cluster is the sum of all exponents of the $\hat{T}_i$ terms within the selected cluster, and a particular term within this expansion is termed a `combination'. For instance in our previous CCSD calculation our size 2 clusters contain the combinations $\hat{T}_1^2$, $\hat{T}_2^2$ and $\hat{T}_1\hat{T}_2$.

As all clusters within a given combination will result in the same excitation level of cluster we can simply only sample those combinations that result in clusters that can couple to a stored coefficient, i.e. those those with excitation level $\le l + 2$. This simplifies \eqref{eq:full-ccsd-combo-expansion} to

\begin{align}
\label{eq:trunc-ccsd-combo-expansion}
\langle D_{\textbf{i}} | N_0e^{\hat{T}}|D_0\rangle = & \langle D_{\textbf{i}} | \left[N_0 \phantom{\hat{T}_2^4}\right. \\
& + \hat{T}_1 + \hat{T}_2 \nonumber \\
& + \frac{1}{2N_0} \left(\hat{T}_1^2 + 2\hat{T}_1\hat{T}_2 + \hat{T}_2^2  \right) \nonumber \\
& + \frac{1}{6 N_0^2} \left(\hat{T}_1^3 + 3 \hat{T}_1^2 T_2\right) \nonumber \\
& + \frac{1}{24 N_0 ^3}  \hat{T}_1^4 \nonumber \\
& \hspace*{50pt} \left.\phantom{\hat{T}_2^4}\right] |D_0\rangle. \nonumber
\end{align}
This is a much simpler form for our expansion, and the benefit of this approach will increase when considering higher levels of coupled cluster theory.

For ease of expression we now define $\epsilon_s$ as the number of possible combinations giving clusters of size $s$, $\eta_{skj}$ as the number of excitors at excitation level $j$ included in combination $k$ at cluster size $s$ and $L_j$ as the summed absolute magnitude of excip populations on excitors of excitation level $j$. $\eta_{skj}$ and $\epsilon_s$ can be precomputed once the truncation level, $l$, of the calculation is known, but $L_j$ varies during the calculation.

While a general form for $\epsilon_s$ is non-trivial to obtain in either the full (as Eq. \eqref{eq:full-ccsd-combo-expansion}) or truncated (as Eq. \eqref{eq:trunc-ccsd-combo-expansion}) expansions numerical results are shown in Table \ref{tab:ncombos-w-tl}. The full expansion shows a rapid increase in the total number of combinations with $l$ due to increasing cluster size and number of different $\hat{T}_i$, while the truncated expansion remains manageable even at truncation level 10. We also highlight the explosion in value of $\epsilon_{l+2}$ in the full expansion, given that only a single combination ever need be selected from this cluster size.

\begin{table}
\begin{tabular}{| c | c | c | c |}
\hline
$l$ & \multicolumn{2}{c |}{$\sum\limits_{i=2}^{l+2}\epsilon_i$} & $\epsilon_{l+2}$ \\ 
\cline{2-3}
 & truncated & full & full \\ [0.5ex]
\hline
2 & 6 & 12 & 5 \\
3 & 12 & 52 & 21 \\
4 & 22 & 205 & 84 \\
5 & 36 & 786 & 330 \\
6 & 57 & 2996 & 1287 \\
7 & 86 & 11,432 & 5005 \\
8 & 127 & 43,749 & 19,448 \\
9 & 182 & 167,950 & 75,582 \\
10 & 258 & 646,635 & 293,930 \\
\hline
\end{tabular}
\caption{Scaling of the number of combinations within the expansion of a coupled cluster wavefunction truncated at $l^{th}$ excitations. \textit{`full'} corresponds to a naive expansion to $\mathcal{O}(\hat{T}^{l+2})$ and \textit{`truncated'} to an expansion including only combinations of excitation level $\le l+2$. The number of combinations of the largest allowed cluster size ($l+2$) within the full expansion is also presented, bearing in mind that only a single combination of this size can affect our calculations ($\hat{T}_1^{l+2}$). It can be seen that the truncated approach results in a drastic reduction in the number of combinations sampled such that it remains manageable even at very high truncation levels.
}
\label{tab:ncombos-w-tl}
\end{table}

These results affirm that the hugely increasing costs of sampling larger clusters when performing even selection can be avoided by constructing an appropriate algorithm.

It should additionally be noted that the term we are sampling within \eqref{eq:new-cc-expansion} can be fully specified by cluster size $s$ and combination number $k$ within a given indexing scheme. Specifically combination $(s,k)$ corresponds to the term $N_0\left(\prod_{j=1}^{l}  \frac{\hat{T}_j^{\eta_{skj}}}{\eta_{skj}!N_0^{\eta_{skj}}}\right)|D_0\rangle$.

We will now select a cluster containing the combination of excitors $e$ as follows.

\begin{enumerate}
\item Select a cluster size $s$ with probability $p_\textrm{size}(s)$ using a to-be-determined distribution.
\item Select a combination $c$ that gives a cluster of size $s$ with probability $p_{\textrm{combo}}(c|s)$ using a to-be-determined distribution. This excludes any combinations resulting in clusters with excitation level greater than $l+2$.
\item Select the appropriate number of excitors from each excitation level, creating a specific combination of excitors $e$ with probability $p_{\textrm{excitors}}(e|c,s)$.
\end{enumerate}

As such $p_{\textrm{select}}(e) = p_{\textrm{size}}(s)p_{\textrm{combo}}(c|s)p_{\textrm{excitors}}(e|c,s)$. We will still wish to apply the lessons learned within section \ref{sec:even-selection} and so to satisfy $\frac{w_e}{n_ap_{\textrm{select}}(e)} = 1$.

If we consider step 3 first, we can see that ensuring $\frac{w_e}{n_{\textrm{a}}p_{\textrm{select}}(e)}$ is the same for all clusters of a given size and combination will require

\begin{equation}
p_{\textrm{excitors}}(e|c,s)\propto \prod\limits_{i=1}^{s} |N_i|.
\label{eq:pexcitors-proportionality}
\end{equation}
Ensuring normalisation in each case gives

\begin{equation}
p_{\textrm{excitors}}(e|c,s) =  \prod\limits_{j=1}^{l} \left( \eta_{scj}!\prod\limits_{i=1}^{\eta_{scj}} \frac{|N_i|}{L_j}\right).
\end{equation}
For clarity the product over $j$ contains all stored coefficient excitation levels (i.e. all $\hat{T}_i$), while that over $i$ is over all selected excitors at the given excitation level $j$ in the cluster. The factor of $\eta_{scj}!$ can again be considered to result from either the number of possible ways to select the same excitors in a different order within an excitation level or the expansion \eqref{eq:new-cc-expansion}.

The constant of proportionality compared to \eqref{eq:pexcitors-proportionality}, $\prod\limits_{j=1}^{l}  \frac{\eta_{scj}!}{L_j^{\eta_{scj}}}$, differs between combinations of the same size. From \eqref{eq:w-expression} we can see this will result in variation of $\frac{w_e}{p_{\textrm{select}}(e)}$ between different combinations of the same size unless we weight $p_{\textrm{combo}}(c|s)$ accordingly. This requires

\begin{equation}
p_{\textrm{combo}}(c|s) \propto \prod\limits_{j=1}^{l}  \frac{L_j^{\eta_{scj}}}{\eta_{scj}!}.
\end{equation}

Defining the reciprocal of the normalisation constant for a given cluster size $s$ as 

\begin{equation}
W_s = \sum\limits_{c=1}^{\epsilon_s} \left(\prod\limits_{j=1}^{l} \frac{L_j^{\eta_{scj}}}{\eta_{scj}!}\right)
\label{eq:normconst-trunc-select}
\end{equation}
we obtain

\begin{equation}
p_{\textrm{combo}}(c|s) = \frac{1}{W_s} \prod\limits_{j=1}^{l}  \frac{L_j^{\eta_{scj}}}{\eta_{scj}!}.
\end{equation}

Combining our current expressions together results in the form

\begin{align}
p_{\textrm{select}}(e) &= p_{\textrm{size}}(s)p_{\textrm{combo}}(c|s)p_{\textrm{excitors}}(e|c,s)\\
		&= \frac{p_{\textrm{size}}(s)}{W_s} \prod\limits_{i=1}^{s} |N_i|.
\end{align}
Finally applying our condition for even selection we obtain the overall expression

\begin{equation}
n_{\textrm{a}}p_{\textrm{size}}(s) = \frac{W_s}{N_0^{s-1}},
\label{eq:nattempts-trunc-select}
\end{equation}
and thus we require

\begin{equation}
n_a = \sum\limits_{s=1}^{l+2} \frac{W_s}{N_0^{s-1}}
\end{equation}
and

\begin{equation}
p_{\textrm{size}}(s) = \frac{W_s}{N_0^{s-1}\sum\limits_{t=1}^{l+2} \frac{W_t}{N_0^{t-1}}}.
\end{equation}

We have thus defined expressions for all relevant selection probabilities in terms of these combinations using only the excip distribution between excitation levels and the possible combinations at each cluster size. We can obtain the former each iteration without considerable extra effort and can calculate the latter within calculation initialisation, so can easily calculate the appropriate probabilities with which to sample our representation each iteration.

We can now perform even selection without $n_{\textrm{a}}$ spiralling out of control. A more thorough discussion of the effects of these changes upon the scaling of $n_a$ is contained within the appendix.

\section{Results}
\label{sec:results}

\subsection{Systems to be considered}

From our considerations within section \ref{sec:new-selection} we can expect to see previous approaches struggling in calculations with large values of $\frac{N_{ex}}{N_0}$. Thus we predict the greatest difference in calculations with high truncation levels, large system sizes and significant multireference character within the wavefunction. We will bear this in mind when choosing our test systems.

Stretched \ce{N2} has significant multireference character in its stretched geometry and high-level coupled cluster theory has previously been benchmarked against the Density Matrix Renormalisation Group\citep{Chan2004} at equilibrium and stretched geometries. Treating this system at various geometries with high truncation levels will enable comparisons to observe the effects of multireference character in our wavefunction upon propagation.

To observe the behaviour of our algorithms with increasing system size we will look at chains of well-separated neon atoms. Since we expect this to be a system with a large reference population we can test algorithmic behaviour with increased system size and basis set cardinality.

\subsection{Calculation Efficiency Measures}

We will use three different measures of calculation cost. These are 
\begin{itemize}
\item $N_{\textrm{ex}}$, the total absolute walker population. This defines the granularity of our stochastic coefficient representation, as a higher population reduces the absolute magnitude of the lowest coefficient value at which we can expect continual occupation. It is useful as a measure of our calculation efficiency, but cannot be easily related to concrete calculation properties.
\item$n_{\textrm{states}}$, the number of occupied excitors. This determines the minimum memory cost of a stable calculation with a given algorithm.
\item $n_{\textrm{attempts}}$, the number of selection attempts made per iteration. This determines the minimum computational cost per iteration, and so the cost of continued propagation to reduce the stochastic error bars on our energy estimate.
\end{itemize}

In all cases the value we are interested in is that at the plateau, as this provides a lower bound on their value when accumulating statistics for the various properties of the wavefunction.

\subsection{Calculation Parameters}

When running calculations we must use sensible values of various parameters to obtain comparable results. In our current approach we must define the timestep used and the initial reference population.

\begin{figure}
\includegraphics[width=0.5\textwidth]{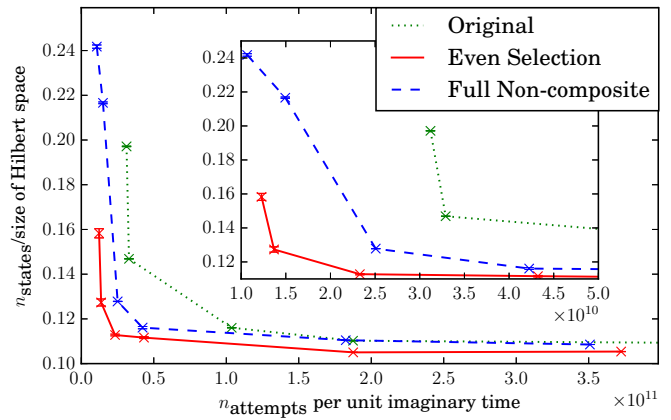}
\caption{Variation in calculation minimum memory costs as a function of the computational cost for all-electron \ce{N2} in a stretched geometry ($r_{\textrm{NN}} =3.6\enskip a_0$) in a Dunning cc-pVTZ basis\citep{DunningJr1989} using CCSDTMC with different selection approaches. Memory costs are given via as the proportion of the truncated Hilbert space occupied at the plateau, while computational costs are given as the equivalent total number of selection attempts required to propagate for a single unit of imaginary time from the plateau. The inset shows details for lower computational cost. Timesteps used were $\delta\tau = 2,4,20,40,100,200\times 10^{-5}$ for even selection, $\delta\tau = 2,4,20,40,200,400\times 10^{-6}$ for full non-composite and $\delta\tau = 0.2,0.4,2,4,20,40\times 10^{-6}$ for the original algorithm. For the original algorithm larger timesteps were unstable.}
\label{fig:timestepdependance}
\end{figure}

The stability of calculations at different timesteps is considered within Figure \ref{fig:timestepdependance}. Within this graph we utilise the number of selections performed per unit time as a measure of computational cost to compensate for the fact that even selection generally has many more attempts per timestep but a correspondingly much larger stable timestep. As the majority of the computational cost of a timestep results from the expensive spawning and death attempts\cite{Vigor2015c} and the number of such attempts is determined by $n_{\textrm{attempts}}$ we can use $n_{\textrm{attempts}}$ to measure of computational cost per iteration. Assuming that the inefficiencies\citep{Vigor2016} of the calculations (adjusted for CCMC) are the same the computational cost of propagating to a given error bar will be proportional to the computational cost per unit imaginary time propagated, which can be calculated as average computational cost per timestep multiplied by the number of timesteps per unit time. The assumption of a uniform inefficiency is sufficient for our current work, but could be a target of future investigation.

From the results in Figure \ref{fig:timestepdependance} we observe that, for any memory cost, even selection results in a less computationally intensive calculation than either previous selection method. We also note that full non-composite selection shows a similar improvement over the original algorithm. With this in mind, for reasons of computational expediency we will from this point onwards only compare between the full non-composite and even+truncated selection approaches. We can also predict from the considerations in section \ref{sec:old-selection} that any computational effects resulting from the under-selection of larger composite clusters in the original algorithm will be comparable in the both algorithms and so be satisfied that we will observe all important calculation behaviour.

Unless otherwise stated, all further calculations use a timestep that results in no bloom events of greater than 3 excips occurring with the even selection algorithm in an equivalent calculation. For the system in Figure \ref{fig:timestepdependance} this gives $\delta\tau = 2 \times 10^{-5}$, and comparison with the other results in Figure \ref{fig:timestepdependance} suggests that this choice of timestep gives a reasonable estimate of the plateau heights that can be expected from a given algorithm while enabling easy determination for a new system. This has the favourable characteristic of being easily determined for a given system without having to converge multiple calculations at different timesteps for each calculation, though provides only a lower bound on the stable timestep with even selection.

We will not consider the variation of $n_{\textrm{attempts}}$ any further within this paper, leaving this instead for later work. For now we only note that the value $\frac{n_{\textrm{attempts}}}{N_{\textrm{ex}}}$, assumed to be 1 originally and 2 in full non-composite sampling, is found to vary during a given even selection calculation and its behaviour is highly system-dependent.

It should be noted that when using full non-composite selection no such general timestep definition is possible, due to the large, but system-dependent, blooms previously noted in Section \ref{sec:justify-new-approach}. This has made finding a stable timestep an exercise in trial-and-error. Being able to apply a single simple criterion to the timestep and obtain a stable calculation is a not inconsiderable benefit when attempting to perform a new calculation.

The choice of initial reference population is less clear. Previous work\cite{Spencer2015a} has noted the dependence of the plateau obtained on the initial population, but for now we will just note that the value of $N_0$ at the plateau is approximately proportional to the initial population, and so a larger initial population gives a smaller peak value of $\frac{N_{\textrm{ex}}}{N_0}$. In full non-composite selection this will change the expected size of blooms at the plateau (and in even selection the plateau $n_{\textrm{attempts}}$ value), though the precise dependence will be strongly nonlinear. As such, we will adopt a convention of starting with 500 excips on the reference and deviating from this when a larger value is required with a comment to this effect. However it should be noted that finding a stable initial population for a given calculation is currently a non-trivial task.

\subsection{Calculation Stability}

\begin{figure}
	{\centering
		\includegraphics[width=0.45\textwidth]{bigbloom.eps}
		\caption*{(a)}
		\includegraphics[width=0.45\textwidth]{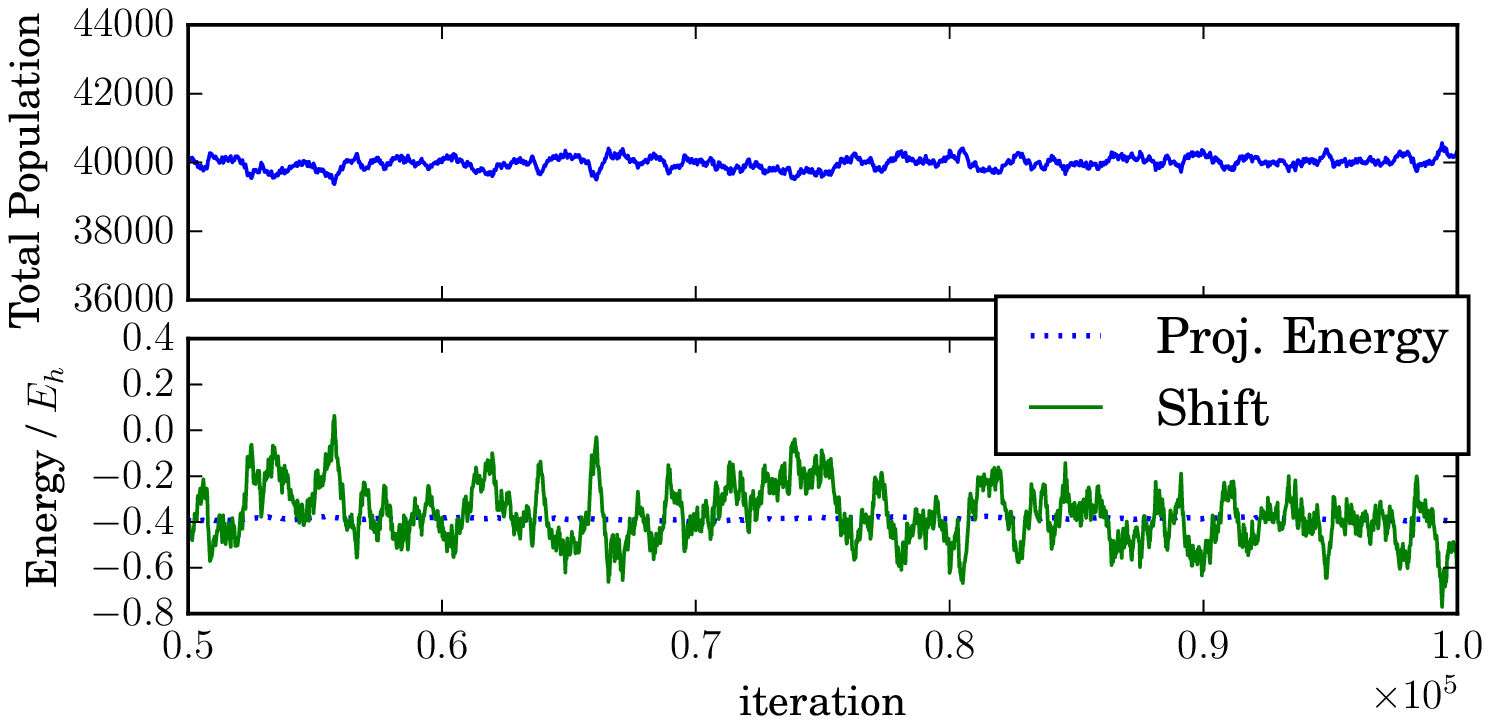}
		\caption*{(b)}\par
	}
	\caption{Calculation dynamics for (CCSDTQ)MC calculations on two well-separated neon atoms in a Dunning cc-pVDZ basis.\citep{DunningJr1989} In each case the top panel shows the total population and the bottom panel shows the energy estimators. (a) shows full non-composite propagation, reproducing Figure \ref{fig:bigbloom_calculation}, and (b) shows the same calculation with the even selection propagation defined in Section \ref{sec:new-selection} for comparison of stability and values.}
	\label{fig:2ne_calculation_comparison}
\end{figure}

\begin{figure}
	{\centering
		\includegraphics[width=0.45\textwidth]{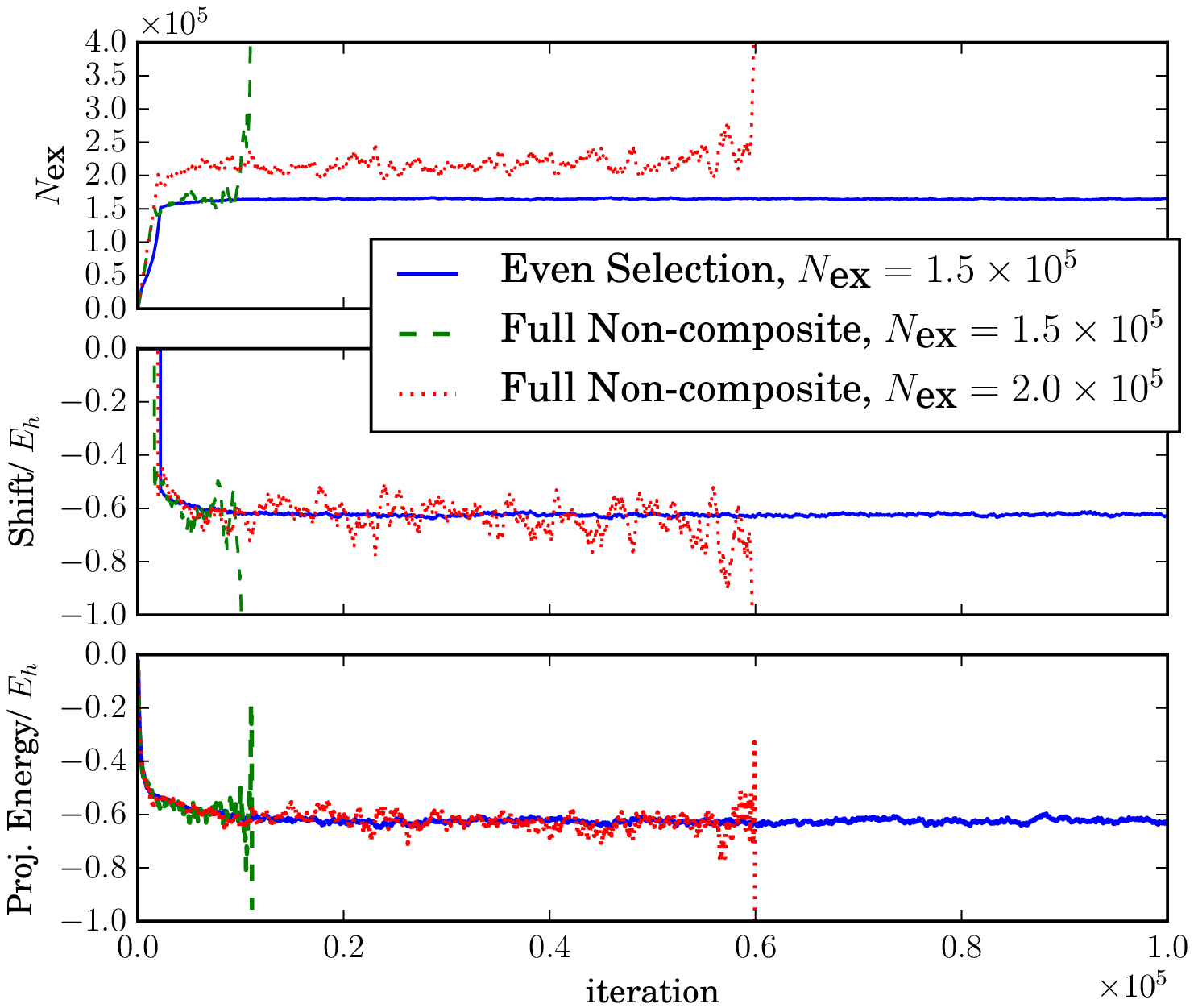}
	}
	\caption{Calculation dynamics for (CCSDT)MC calculations on frozen-core stretched \ce{N2} ($r_{\textrm{NN}}=3.6\textrm{a}_{\textrm{0}}$) in a Dunning cc-pVDZ basis.\citep{DunningJr1989} The top panel shows total excip population, the middle panel the population control parameter $S$ and the bottom the instantaneous projected energy. Calculations all use $\delta\tau=0.0023$ and $N_0=500$, and are labelled with the selection algorithm used and the target population at which population control was initiated.}
	\label{fig:stretched_n2_calculation_comparison}
\end{figure}

First and foremost, we compare calculation stability with the modified algorithm to previous approaches. We can see in Figures \ref{fig:2ne_calculation_comparison} and \ref{fig:stretched_n2_calculation_comparison} that the new selection algorithm avoids the stability issues previously present, stabilising excip population and energy estimates within a much smaller range during a calculation. The bloom events observed within Figure \ref{fig:bigbloom_calculation} are entirely absent, as can be seen from the comparison within Figure \ref{fig:2ne_calculation_comparison}.

Within Figure \ref{fig:stretched_n2_calculation_comparison} we can also see the effect of increasing excip population at which statistics are collected upon full non-composite calculation stability. At both target populations the calculation initially appears stable under population control before spontaneously destabilising  due to a series of large blooms at $\tau=23$ and 138 units of imaginary time respectively. This demonstrates that while a higher target population appears to give a more stable calculation for a time, this is no guarantee that we can reliably continue to extract statistics indefinitely.

In comparison the even selection algorithm should be possible to propagate indefinitely without any such destabilisation, and in this case saw no major variations over $10^5$ iterations ($\tau=230$).

\subsection{\ce{N2}}

\begin{figure}
\includegraphics[width=0.5\textwidth]{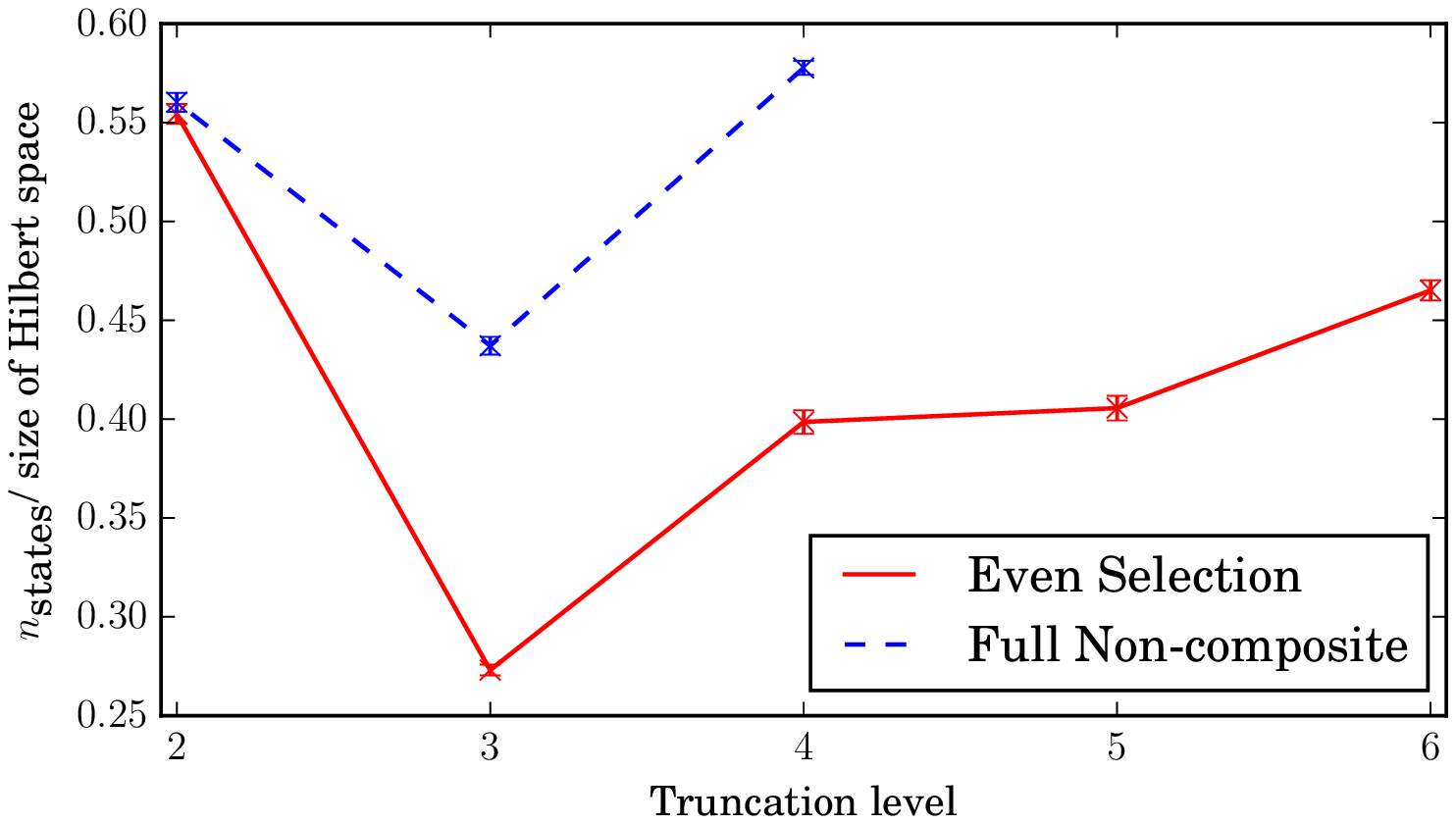}
\caption{Memory cost as a function of CC truncation level for frozen-core \ce{N2} in an equilibrium geometry ($r_{\textrm{NN}}=2.118\textrm{a}_{\textrm{0}}$) and a cc-pVDZ basis\citep{DunningJr1989} when using different selection approaches. Full non-composite propagation was not found to give a stable calculation for truncation levels 5 or 6 with any initial parameter combinations previously defined.}
\label{fig:n2-2.118-nstates}
\end{figure}

\begin{figure}
\includegraphics[width=0.5\textwidth]{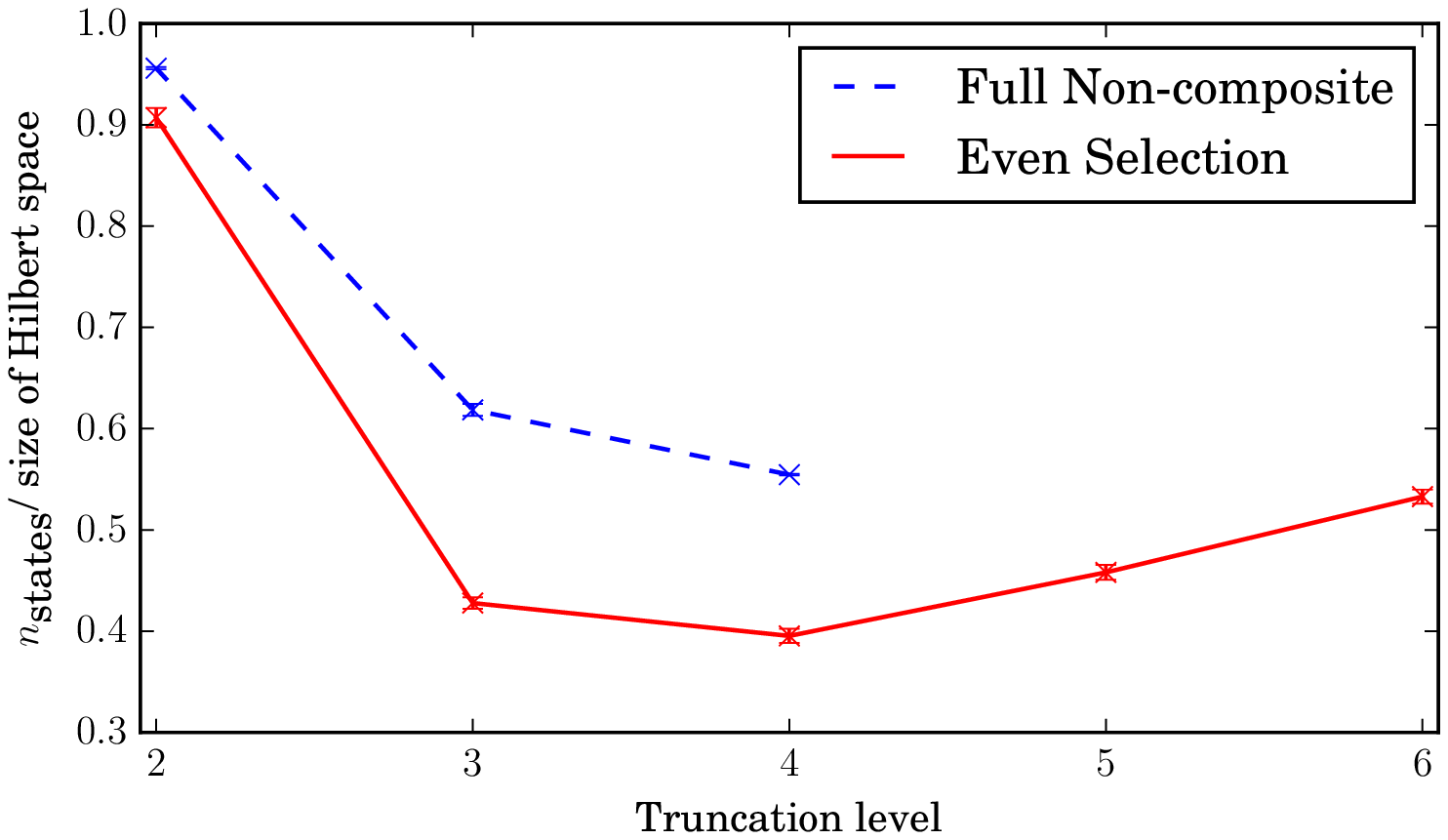}
\caption{Memory cost as a function of CC truncation level for frozen-core \ce{N2} in a stretched geometry ($r_{\textrm{NN}}=3.6\textrm{a}_{\textrm{0}}$) and a cc-pVDZ basis\citep{DunningJr1989} when using different selection approaches. Full non-composite propagation was not found to give a stable calculation for truncation levels 5 or 6 with any initial parameter combinations previously defined, while truncation level 4 required a timestep a factor of 10 smaller than that already defined.}
\label{fig:n2-3.6-nstates}
\end{figure}

Results for the stretched and equilibrium geometries of \ce{N2} at a range of truncation levels are presented in Figures \ref{fig:n2-2.118-nstates} and \ref{fig:n2-3.6-nstates}. These results utilise frozen core electrons and a Dunning cc-pVDZ basis.\citep{DunningJr1989}

In all cases the new selection algorithm is found to have a lower memory cost than the previous approach, while also providing a stable calculation at all truncation levels with the previously specified parameter set in both geometries. In comparison, the full non-composite algorithm required timestep reduction by a factor of 10 to be stable for CCSDTQ in the stretched geometry and was not stable at higher truncation levels for either geometry regardless of initial population or timestep. This suggests that, as expected, the full non-composite algorithm struggles at higher truncation levels in systems with significant multireference character in the wavefunction.

It should also be noted that the increase in the proportion of the Hilbert space occupied in both these systems as truncation levels approach CCSDTQ56 is likely the result of significant multireference character entering the wavefunction. In this case a sextuple excitation of the reference will be highly-weighted so we can expect significant contributions around this excitation level. From this we would expect the memory cost to decrease at even higher truncation levels, but this is left for a further study to confirm.

\subsection{Ne}

\begin{table}
\begin{tabular}{c c c c c c}

\hline\hline
Basis & CC & & $\textrm{n}_{\textrm{states}}$& \\ 
Set & Level & Even & Full & Hilbert \\ 
& & selection & non-composite & Space \\ [0.5ex]
\hline
cc-pVDZ & 2 & 180 & 190 & 398(1) \\
& 3 & 360 & 347 & 4677(9) \\
& 4 & 825(9) & 850(20) & 3065(5) \\
& 5 & 1380(10)& 1589(14) & 1.135(2)$\times 10^5$ \\
& 6 & 2060(30) & 2325(60) & 2.595(3)$\times 10^5$ \\
cc-pVTZ & 2 & 1.90(2)$\times 10^3$ & 1.94(1)$\times 10^3$ & 2.957(7)$\times 10^3$ \\
& 3 & 9.7(1)$\times 10^3$ & 9.39(3)$\times 10^3$ & 1.030(2)$\times 10^5$ \\
& 4 & 5.36(2)$\times 10^4$ & 5.73(6)$\times 10^4$ & 1.967(2)$\times 10^6$\\
& 5 & 2.71(7)$\times 10^5$ & 2.84(4)$\times 10^5$ & 2.124(2)$\times 10^7$\\
& 6 & 1.44(4) $\times 10^6$ & 1.620(9)$\times 10^6$ & 1.364(1)$\times 10^8$\\
cc-pVQZ & 2 & 4.78(1)$\times 10^3$ & 4.79(2)$\times 10^3$ & 1.157(2)$\times 10^4$ \\
& 3 & 3.200(2)$\times 10^4$ & 3.257(8)$\times 10^4$ & 8.30(1)$\times 10^5$ \\
& 4 & 3.745(3) $\times 10^5$& 3.830(4)$\times 10^5$ & 3.210(3)$\times 10^7$ \\
& 5 & 7.207(3)$\times 10^6$ & 8.01(3)$\times 10^6$ & 7.047(9)$\times 10^8$ \\
\end{tabular}
\caption{Behaviour of calculation memory costs for a single neon atom at different basis sets and truncation levels. The even selection and full non-composite columns contain the memory cost at the plateau, while the full reduced Hilbert space (as would be stored in convention coupled cluster theory) is provided for comparison.}
\label{tab:neon-basis-results}
\end{table}

\begin{figure}
\includegraphics[width=0.5\textwidth]{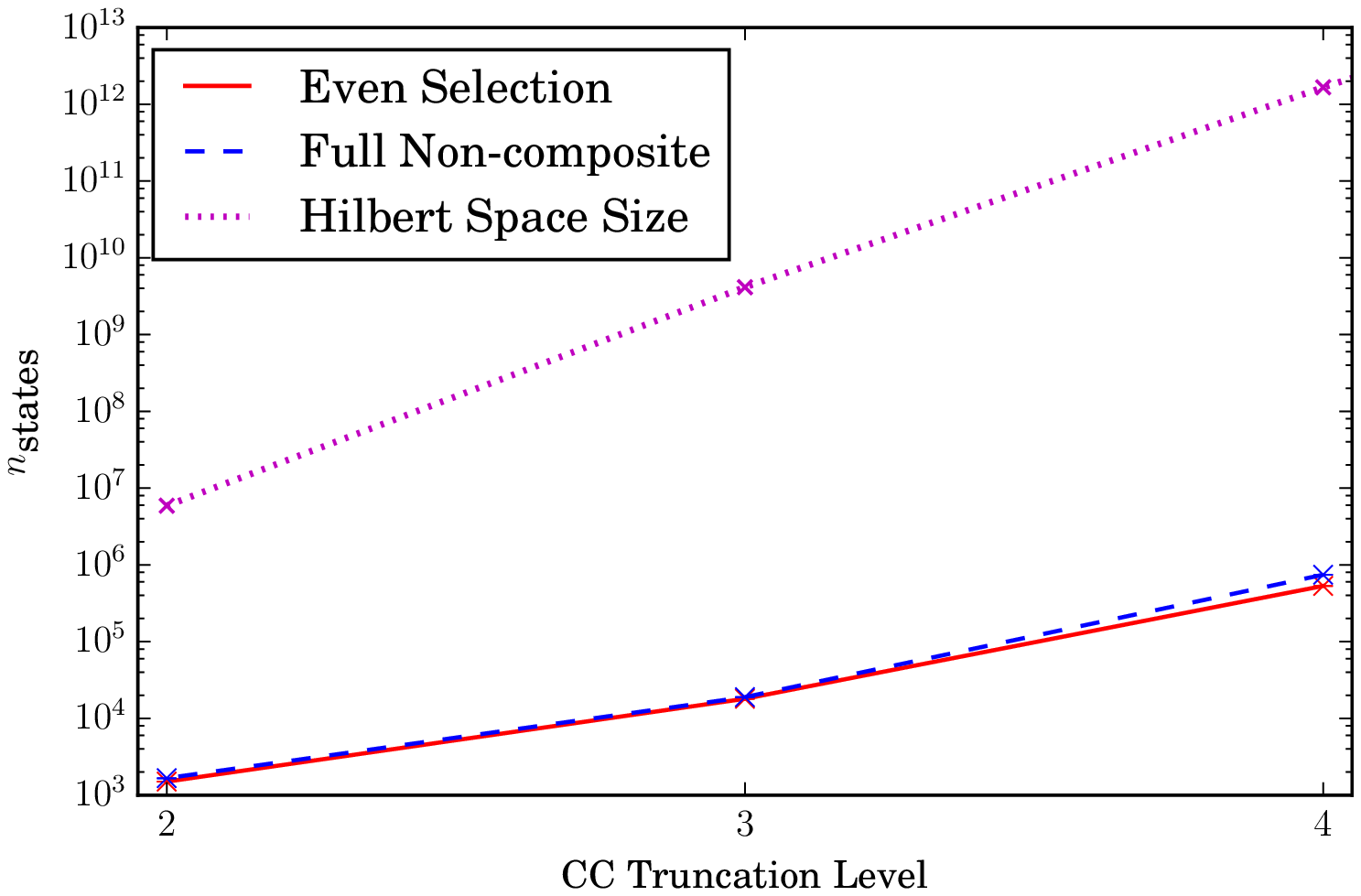}
\caption{Number of excitors in the stochastic CCMC wavefunction at the plateau compared to the full Hilbert space for 4 well-separated Ne atoms. Orbitals were localised according to the Foster Boys criterion.\citep{Boys1960} We can see that in this system both selection algorithms perform comparably, and are capable of performing a stochastic CCSDTQ calculation with significantly lower memory costs than conventional CCSD in the same system.}
\label{fig:4ne-nstates}
\end{figure}

\begin{figure}
\includegraphics[width=0.5\textwidth]{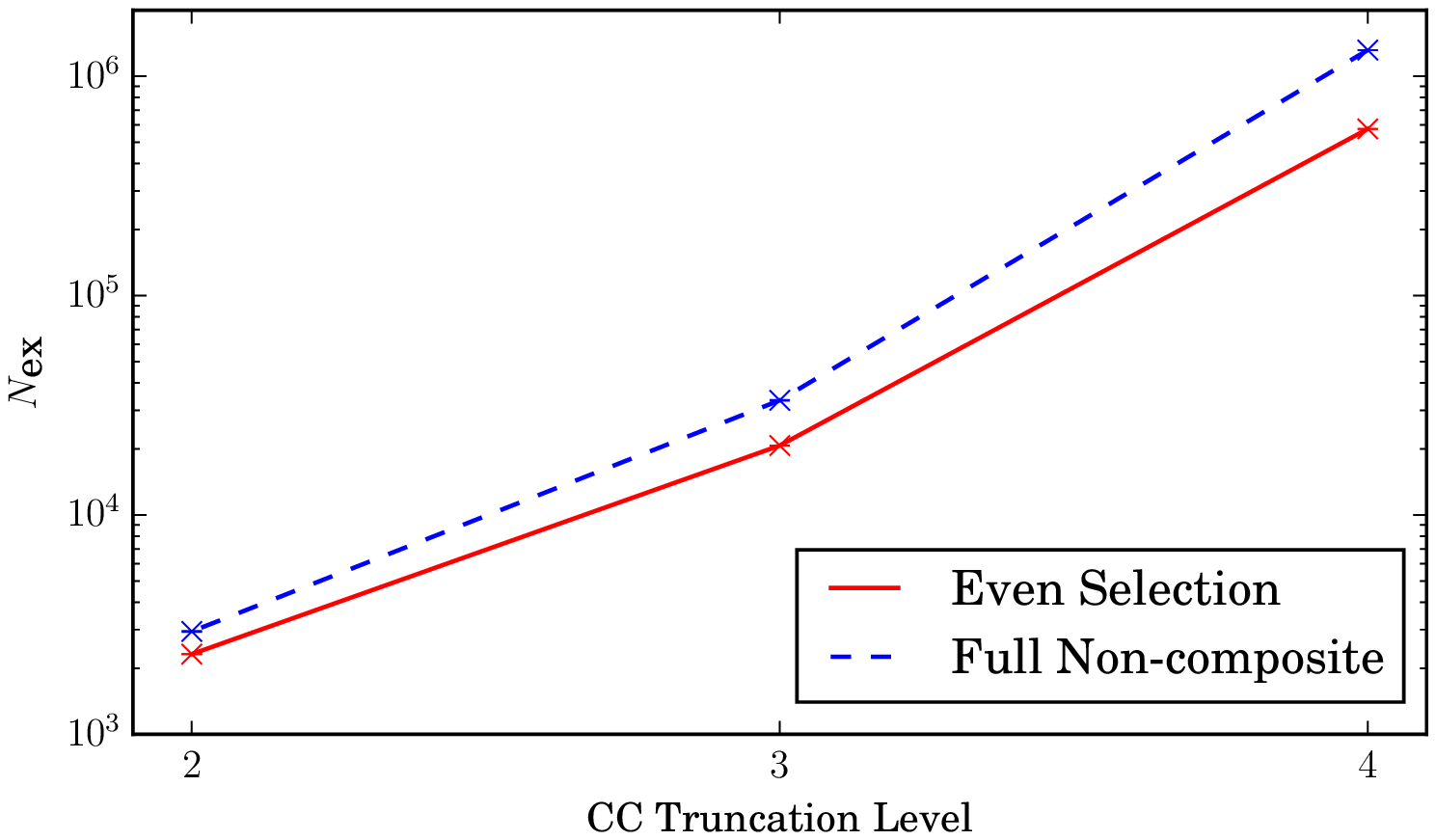}
\caption{Number of excips in the stochastic CCMC wavefunction at the plateau for 4 well-separated Ne atoms. Orbitals were localised according to the Foster Boys criterion.\citep{Boys1960} We can see that in this system using even selection results in a smaller excip population, a pattern that is repeated in most systems.}
\label{fig:4ne-nexcips}
\end{figure}

Results for the application of full non-composite and even selection algorithms to a neon atom as we increase basis set size are presented in Table \ref{tab:neon-basis-results}. It can be seen that for a cc-pVDZ basis set and truncation level the conventional selection approach behaves comparatively to even selection, but as we increase basis set cardinality even selection provides a more tangibly reduced memory cost.

Similar results for chains of well-separated ($r$=$1000 a_{0}$) neon atoms are presented in Table \ref{tab:neon-size-results} and plotted in Figure \ref{fig:4ne-nstates} for a chain of 4 neon atoms. By utilising orbitals localised according to the Foster-Boys criterion\citep{Boys1960} it can be seen that stochastic CCMC provides much reduced memory costs due to its exploitation of the inherent sparsity of the local representation. Utilising even selection reinforces this, giving a further reduced memory cost at higher truncation levels.

A decrease in $N_{\textrm{ex}}$ at the plateau is also observed when using even selection, as demonstrated in Figure \ref{fig:4ne-nexcips}. This shows a reduction in the granularity of our representation required for convergence, as would be expected.

\begin{table}
\begin{tabular}{c c c c c c}
\hline\hline
System & CC & & $\textrm{n}_{\textrm{states}}$& \\ 
 & Level & Even & Full & Hilbert \\
 & & selection & non-composite & Space \\ [0.5ex]
\hline
2 $\times$ Ne & 2 & 672(2) & 615(8) & 4.653$\times 10^4$ \\
& 3 & 2328(3) & 2.80(3)$\times 10^3$ & 2.720 $\times 10^6$ \\
& 4 & 2.691(9)$\times 10^4$ & 3.468(2)$\times 10^4$ & 8.666 $\times 10^7$ \\
& 5 & 3.26(1)$\times 10^5$ & 3.319(2)$\times 10^5$ & 1.670 $\times 10^9$ \\
& 6 & 3.348(3)$\times 10^6$ & n/a & 2.089 $\times 10^{10}$ \\
4 $\times$ Ne & 2 & 1500(1) & 1662(6) & 5.885 $\times 10^6$ \\
& 3 & 1.7992(1)$\times 10^4$ & 1.89(1)$\times 10^4$ & 4.136 $\times 10^9$ \\
& 4 & 5.310(2)$\times 10^5$ & 7.425(2)$\times 10^5$ & 1.664 $\times 10^{12}$ \\
\end{tabular}
\caption{Behaviour of calculation memory costs for chains of well-spaced ($r$=$1000a_{0}$) neon atoms of different length for different CC truncation levels. Calculations used a Dunning cc-pVDZ basis\citep{DunningJr1989} sets localised according to the Foster-Boys criterion. The columns are as in \ref{tab:neon-basis-results}. It can be seen that while both stochastic approaches provide a much reduced memory cost due to the sparsity of the local representation, an improved selection approach is required to fully utilise this sparsity at higher truncation levels.}
\label{tab:neon-size-results}
\end{table}

\section{Conclusions}

In this paper we have demonstrated that by reconsidering the sampling of our wavefunction representation within stochastic coupled cluster theory we can obtain an algorithm with lower memory and computational costs as well as providing much more stable calculations. The benefits of this approach are particularly evident at larger truncation levels, system sizes and basis sets, as well as in relatively multireference systems.

This new algorithm entirely removes uncontrollable blooms from CCMC calculations, avoiding the potential issues associated with this behaviour. This includes spontaneous destabilisation of calculations and causing stochastic error bars on energy estimates to become unreliable, both of which are cause for serious concern.

The removal of blooms from calculations is also vital to various approaches for future development within stochastic coupled cluster theory. Let us, for instance, consider an MPI-based parallelisation scheme for stochastic coupled cluster theory that somehow divides excitors between processes and samples in such a way as to approximate the undivided sampling, analogous to the approach previously described for FCIQMC.\cite{Booth2014} If the expansion contains instantaneously extremely heavily-weighted excitors, as results from large blooms, then either the processes on which these excitors are located must perform an increased number of samples or clusters containing these excitors must be undersampled  due to their greater weight within the expansion. The former approach will result in very poor load balancing, while the latter will exacerbate the underlying problem of the blooms themselves. Such a parallelisation scheme has already been implemented within HANDE\cite{Spencer2015} by Spencer et al and will be detailed in a forthcoming paper. Adaptation of even selection to this scheme is thus essential to exploiting modern computing resources.

Removing the requirement to determine calculation parameters that depend non-linearly on the system considered by providing a rigorous method to define the order of magnitude of the timestep is a significant development towards enabling black-box usage of stochastic coupled cluster theory. Similar approaches for the initial reference population and shift damping are currently in preparation and will enable calculations to be performed with a minimum of predetermined values. Only the maximum allowed memory cost would remain as a parameter to be specified within a calculation.

Following from this work we can consider the impact of our refined sampling upon  the previously considered initiator approximation.\citep{Spencer2015a} This could be expected to provide a large benefit by enabling effective sign-consistent sampling as previously observed to be essential to the convergence of initiator error within \textit{i}FCIQMC.\citep{Cleland2010}

\section{Acknowledgements}

C.J.C.S. is grateful to the Sims Fund for a studentship and A.J.W.T. to the Royal Society for a University Research Fellowship under Grant No. UF110161. Both are grateful for support under ARCHER Leadership Project grant e507.

Molecular Orbital integrals for all systems were generated using PySCF\citep{Sun2017} and all post-Hartree Fock calculations were performed using a development version of HANDE.\citep{Spencer2015} Figures were plotted using matplotlib.\cite{Hunter2007} Raw and analysed data and analysis scripts are freely available online.\cite{Scott}


\section*{Appendix: The Scaling of Sampling Attempts With and Without Truncation}

If we assume that the average number of excips per possible excitor is a constant value between given calculations and excitation levels we obtain

\begin{equation}
L_{j} \propto N_{\textrm{excitors}}(j)
\end{equation}
and

\begin{equation}
N_{\textrm{excips}}^{(l)} = \sum\limits_{j=1}^{l}L_{j} \propto \sum\limits_{j=1}^{l} N_{\textrm{excitors}}(j),
\end{equation}
where $N_{\textrm{excips}}^{(l)}$ is the number of excips required for a calculation at truncation level $l$ and $N_{\textrm{excitors}}(j)$ is the number of possible excitors of excitation level $j$ (i.e. that would contribute to $\hat{T}_j$ in non-stochastic Coupled Cluster). Note that all these expressions explicitly exclude the reference and its excip population.

Considering all possible combinations of occupied and virtual orbitals for a system with $n_e$ electrons and $M$ basis functions gives the number of such excitors as

\begin{equation}
N_{\textrm{excitors}}(j) = \phantom{}^{M}C_{j} \phantom{}^{n_e}C_{j},
\end{equation}
which for $j<<n_e<M$ gives the scaling

\begin{equation}
\mathcal{O}(N_{\textrm{excitors}}(j)) = \mathcal{O}(M^jn_e^j).
\end{equation}

If both $n_e$ and $M$ are parametrised by an arbitrary function of a system size $N$ we obtain

\begin{equation}
\mathcal{O}(N_{\textrm{excitors}}(j)) = \mathcal{O}(N^{2j})
\end{equation}
and so

\begin{equation}
\mathcal{O}(N_{\textrm{excips}}^{(l)}) = \sum\limits_{j=0}^{l} \mathcal{O}(N^{2j}) \approx \mathcal{O}(N^{2l}).
\end{equation}

We note that since $n_{a}$ was previously fixed to be linear in $N_\textrm{excips}$ this will also be the hypothetical scaling of $n_a$ within the methods discussed in section \ref{sec:old-selection} if propagation with these approaches is stable. This is termed hypothetical since increasing bloom sizes with system size would actually lead to an increase in $N_{\textrm{excips}}^{(l)}$.

Substituting this scaling into Eq. \eqref{eq:nattempts-pure-even} gives us a simple expression for the scaling of $n_a$ with system size if we used only even selection as initially proposed in section \ref{sec:even-selection} results in

\begin{equation}
\mathcal{O}(n_a^{(l)}) = \mathcal{O}(N_{\textrm{excips}}^{(l)}) \sum\limits_{s=0}^{l+2} \frac{1}{s!}\left(\frac{\mathcal{O}(N_{\textrm{excips}}^{(l)})}{\mathcal{O}(N_0)}\right)^{s-1}.
\end{equation}
Assuming that $N_0$ is a constant value as system size increases and taking the largest term within the sum gives

\begin{align}
\mathcal{O}(n_a^{(l)}) &= \mathcal{O}(N_{\textrm{excips}}^{(l)})^{l+2} \\
&= \mathcal{O}(N^{2l(l+1)}).
\label{eq:scaling-init-even-select}
\end{align}

To obtain a similar expression when using truncated selection as in section \ref{sec:truncated-selection} using \eqref{eq:nattempts-trunc-select} is more involved, as it is dependent upon the allowed cluster combinations (as defined in the main text).

This task can be simplified by the observation that since we have implicitly assumed all excitors are occupied selecting a cluster is equivalent to selecting the constituent second-quantised creation and annihilation operators. Thus selection of a cluster of excitation level $j$ involves selecting $2j$ second-quantised operators, and as such will scale as $\mathcal{O}(N^{2j})$. This gives the scaling of $n_a$ with system size when using truncated selection as

\begin{equation}
\mathcal{O}(n_a^{(l)}) = \mathcal{O}(N^{2(l+2)}).
\label{eq:scaling-truncated-select}
\end{equation}

Comparison of Eqs. \eqref{eq:scaling-init-even-select} and \eqref{eq:scaling-truncated-select} demonstrates the requirement to use truncated selection if one is to avoid calculation costs rapidly becoming totally untenable.

It should be noted that the assumptions made in this derivation represent the worst-case scenario and so an upper bound on the scaling of calculations. We would expect the average excip population per excitor to fall rapidly with truncation level, system size and basis set due to a smaller proportion of the possible excitors being required in the stochastic wavefunction representation, as seen in Fig. \ref{fig:4ne-nstates} and Table \ref{tab:neon-basis-results}.

\bibliography{./bibliography}

\end{document}